\documentclass[%
 reprint,
 amsmath,amssymb,
 aps,
]{revtex4-1}

\usepackage{graphicx}% Include figure files
\usepackage{dcolumn}% Align table columns on decimal point
\usepackage{bm}% bold math

\usepackage{lineno,hyperref}
\usepackage{epsfig}
\usepackage{float}
\usepackage{bookmark}
\usepackage{times}
\usepackage{amsmath}
\usepackage{color}
\usepackage{epstopdf}
\allowdisplaybreaks[4]
\usepackage{stfloats}

\begin{document}

\preprint{APS/123-QED}

\title{Game-theoretical approach for opinion dynamics on social networks}

\author{Zhifang Li}

\author{Xiaojie Chen}
\email{xiaojiechen@uestc.edu.cn (corresponding author)}
\affiliation{School of Mathematical Sciences, University of Electronic Science and Technology of China, Chengdu 611731, China}

\author{Han-Xin Yang}
\affiliation{Department of Physics, Fuzhou University, Fuzhou 350108, China}
\author{Attila Szolnoki}
\affiliation{Institute of Technical Physics and Materials Science, Centre for Energy Research, P.O. Box 49, H-1525 Budapest, Hungary}

\begin{abstract}
Opinion dynamics on social networks have been received considerable attentions in recent years. Nevertheless, just a few works have theoretically analyzed the condition in which a certain opinion can spread in the whole structured population. In this paper, we propose an evolutionary game approach for a binary opinion model to explore the conditions for an opinion's spreading. Inspired by real-life observations, we assume that an agent's choice to select an opinion is not random, but is based on a score rooted both from public knowledge and the interactions with neighbors. By means of coalescing random walks, we obtain a condition in which opinion $A$ can be favored to spread on social networks in the weak selection limit. We find that the successfully spreading condition of opinion $A$ is closely related to the basic scores of binary opinions, the feedback scores on opinion interactions, and the structural parameters including the edge weights, the weighted degrees of vertices, and the average degree of the network. In particular, when individuals adjust their opinions based solely on the public information, the vitality of opinion $A$ depends exclusively on the difference of basic scores of $A$ and $B$. When there are no negative (positive) feedback interactions between connected individuals, we find that the success of opinion $A$ depends on the ratio of the obtained positive (negative) feedback scores of competing opinions. To complete our study, we perform computer simulations on fully-connected, small-world, and scale-free networks, respectively, which support and confirm our theoretical findings.
\end{abstract}

\maketitle

\textbf{The spreading of opinions on social networks can be detected in several ways in modern times and to understand related collective behaviors is the focus of research interest in last decades. Several examples can be raised ranging from political elections to fashion and marketing which influence our daily life. Albeit the microscopic process seems to be simple, it is still challenging to find analytical solutions. Our present theoretical approach utilizes the fact that to change an opinion, which can be considered as a decision-making process, depends not only public information, but also on local interactions. In particular, we introduce an evolutionary game-theoretical approach with which an opinion update depends directly on a score. This score is calculated both from globally reachable public knowledge and also from restricted interactions with neighbors. Our principal goal is to provide theoretical conditions which ensure the spreading of opinion $A$ in a binary opinion model. By means of coalescing random walks, we obtain a theoretical condition and study different cases to identify decisive factors. When only public information is available, the final evolutionary outcome depends exclusively on the difference of basic scores of binary opinions. However, when only local interactions without negative (positive) feedbacks with neighbors are considered, an opinion's final diffusion depends sensitively on the ratio of the obtained positive (negative) feedback scores of competing opinions. For the successful spreading of the opinion this ratio should exceed a critical value, which depends on the degree values of nodes on unweighted networks. For a more comprehensive study, we have checked interaction graphs with different topologies and confirmed our theoretical predictions by computer simulations.}

\section{Introduction}
Because of its paramount importance in several social phenomena, the study of opinion dynamics in structured populations has become an intensively studied research area in the last decades
~\cite{Proskurnikov2017ARC,Acemoglu2011DGA,Lin2018IEEE TAC,Proskurnikov2015IEEE TAC,Ghaderi2014Automatica,Yildiz2013ACMTEC,Battiston2020PR}. In order to study the evolution and diffusion of opinions among interacting agents, a huge variety of mathematical models have been proposed~\cite{Castellano2009RMP,DeGroot1974JASA,Friedkin1990JMS,Weisbuch2004EPJB,Holcombe1998PC,Galam2002EPJBMCS,Sznajd-Weron2000IJMPC,Deffuant2000ACS,Krause2002,Hegselmann2005CEF}. In general, these models can be divided into two main categories: the first branch assumes discrete opinion model in which individuals take the discrete opinion values, like voter model~\cite{Holcombe1998PC}, majority rule model~\cite{Galam2002EPJBMCS}, or Sznajd model~\cite{Sznajd-Weron2000IJMPC}. The other approach uses continuous opinion models in which continuously distributed opinion values are considered, including Deffuant model~\cite{Deffuant2000ACS}, HK model (Hegselmann-Krause model)~\cite{Krause2002,Hegselmann2005CEF}, and so on.

It is worth mentioning that the above mentioned models have provided theoretical paradigms for studying opinion dynamics on social networks. In
a realistic world, when an individual chooses or changes a certain opinion, his/her choice would be affected by the actions of neighbors. Accordingly there exists opinion interaction between two connected individuals~\cite{Jiang2014IEEETSP,Zhang2021IEEETIFS,Di2007IJMPC}. Generally, individuals tend to make a rational choice and prefer to choose the opinion which can induce positive interactions with neighbors. From this viewpoint, how to choose an opinion can be regarded as a decision-making process with strategic interactions~\cite{Di2007IJMPC}. In general this is a missing feature from previous theories, and thus it is meaningful to characterize the microscopic process of opinion choice in the mentioned way~\cite{Wu2020CCC,Ding2009IJMPC,Ding2010PASMA,Di2007IJMPC,Yang2016EPL,Zhou2018EPL,Zino2020CDC}. Evolutionary game theory, as a powerful mathematical tool, can be used to achieve this goal. Recently, some related efforts have been made along this research path. For example, Di Mare and Latora studied how to use game theory in the modeling of opinion formations in a way to simulate the basic interaction mechanisms between two individuals and particularly have shown how opinion formation can be obtained by just changing the rules of the game~\cite{Di2007IJMPC}. Subsequently, Yang proposed a consensus model of binary opinion in the framework of evolutionary games and studied how the necessary time to reach a consensus state can be reduced~\cite{Yang2016EPL}. It is found that there exists an optimal cost-benefit ratio in the game leading to the shortest consensus time. Furthermore, Zhou \emph{et al}. introduced conformity-driven teaching ability into the evolutionary process of opinion dynamics~\cite{Zhou2018EPL}. By means of computer simulations, they found that when the teaching ability strongly depends on the conformity, the consensus time can be shortened significantly. On the other hand, Zino~\emph{et al}. proposed a novel model that captures the coevolution of actions and opinions on social networks and considers the interplay between the dynamics of actions and opinions. To be more specific, each agent updates his/her opinion, depending on the opinions shared by others, the actions observed on the network, and possibly an external influence source~\cite{Zino2020CDC}. However, the mentioned work does not take into account the microscopic opinion updating with strategic interactions. It is worth pointing out that most of previously related works are based on computer simulations and thus far a few studies have tried analytical calculations to identify the theoretical conditions of successful spreading of opinion in a structured population.

In this work, we thus would like to  analyze the evolutionary process of opinion spreading on social networks theoretically where we integrate the evolutionary game approach. Accordingly, we consider binary opinion and assume that each individual can choose one of the two opinions $A$ and $B$. When an individual chooses an opinion, he/she can obtain a basic score based on the available public information. This may correspond to the phenomena in human society that, for example, when a person wants to choose between two unknown vacation places for a vacation plan. For a better decision, he/she will first make a basic evaluation of the potential destinations based on public information available from tourist brochures in the Internet and public opinions in some social platforms. Based on these public information, he/she will have some different preferences about the two vacation places. Alternatively, one can represent the preference on each option by a basic score value. In addition, the opinion choice of individuals are also influenced by the actions of their neighbors or friends. We therefore also assume that individuals can interact with each others via pairwise interactions~\cite{Traulsen2007JTB}. In particular, when an individual interacts with a neighbor who shares the same opinion, our agent receives a positive score due to a positive supporting feedback effect. Otherwise, our agent gets a negative score because of the conflicting decisions which can be implemented as a negative feedback. This is a psychologically reasonable assumption because the positive score received from an interaction with akin neighbor expresses a sense of belonging to the same group, and the negative score represents a stress of nonconformity~\cite{Rabin1993AER,Cao2008PRE,Szolnoki2015IF,Szolnoki2018NJP}. This argument establishes a certain interaction of neighboring agents that can be handled in the framework of a game-theoretical model~\cite{Tadelis2013,Weibull1997,Hofbauer1998,Sigmund1999CB}. Having considered the mentioned interactions, the final score can be calculated which determines how an agent updates personal opinion. In general, we can say that during the evolutionary process individuals will imitate neighboring opinions which provide higher individual score for them and prefer to choose a neighbor who has a higher level of intimacy with them to follow~\cite{Yang2016EPL,Peng2018ASC}.

Based on the above description, we propose an evolutionary game-theoretical model of binary opinions on social networks and accordingly study the evolution of binary opinions with strategic interactions. Indeed, one crucial quantity for studying evolutionary dynamics of binary opinions on networks is the fixation probability $\rho_A$ of opinion $A$, which means the probability that individuals with opinion $A$ take over the whole structured population given that initially an individual at a vertex is chosen randomly to have opinion $A$ in a population of individuals with opinion $B$. We would like to point out that initially a node in the network is randomly selected to be the seeding node and the opinion is set to $A$. By means of using evolutionary game theory, we derive the formula of fixation probability of opinion $A$ by calculating coalescence times~\cite{Kingman1982SPA,Wakeley2009,Oliveira2012TAMS,Allen17Nature}, and obtain the condition in which opinion $A$ can be favored to spread in the weak selection limit. We find that whether opinion $A$ can diffuse or not depends on the score parameters we considered, the edge weights, the weighted degrees of vertices, and the average degree of the network. Particularly, when individuals only adjust their opinions based on the basic score derived from the information of public knowledge, we find that the success spreading of opinion $A$ depends solely on the difference of basic scores of $A$ and $B$. Besides, we consider the special case of our game model in which the negative (positive) feedback effects of strategic interactions of opinions are ignored and the two basic scores about opinion $A$ and $B$ are the same for the evolution of binary opinions. Then we find that opinion $A$ can be favored in the weak selection limit if the ratio of positive (negative) scores of competing opinions exceeds a critical value. We finally carry out computer simulations on fully-connected, small-world, and scale-free networks to confirm the robustness of our theoretical predictions.

The rest of this paper is structured as follows. We first introduce the basic definition and construct our model in Section $2$. Then, we obtain the condition in which opinion $A$ can be favored to spread under weak selection and carry out simulations to validate our theoretical results in Section $3$. Finally, we summarize our conclusions in Section $4$.

\section{Model}\label{section2}

\begin{figure*}[!t]
\begin{center}
\includegraphics[width=4in]{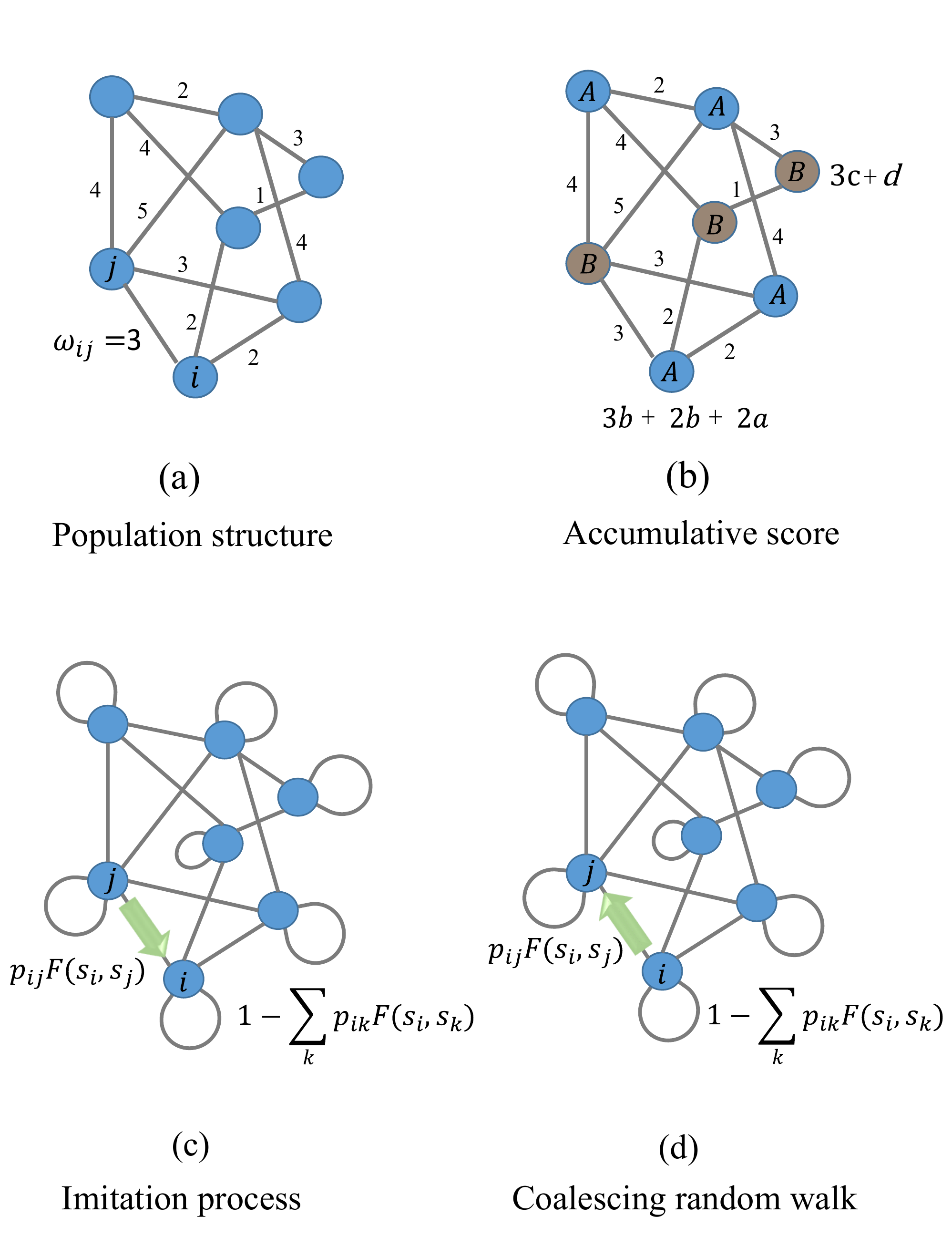}
\caption{Illustration of interaction-based opinion dynamics on weighted graphs. In panel (a), population structure is represented by a weighted and connected graph $G$ with edge weights $\omega_{ij}$. In panel (b), individual $i$ interacts with nearest neighbors and gets an edge-weighted accumulated score in state $\textbf{s}$, denoted by $P_i(\textbf{s})$, which means the sum of scores obtained from each neighbor multiplied by the corresponding edge weight. In panel (c), a new graph $G'$ is generated based on the graph $G$. The green arrow on $G'$ indicates that individual $i$ imitates the opinion of individual $j$. The imitation probability is $p_{ij}F(s_i,s_j)$ for $j\neq i$, otherwise individual $i$ maintains the original
opinion with probability $1-\sum_{k}{p_{ik}F(s_i,s_k)}$, indicated by the self-loop in the figure. In panel (d), individual $i$ performs coalescing random walks on $G'$. The arrow in panel $(d)$ indicates that a step from $i$ to $j$ is taken with probability $p_{ij}F(s_i,s_j)$. Here, coalescing random walk is a process of looking for ancestors backward, which is dual to the neutral case in our model.} \label{figure1}
\end{center}
\end{figure*}

We consider a structured population of individuals with size $N$. The interaction graph is represented by a weighted graph $G$ with edge weight $\omega_{ij}\geq 0$, where individuals are located on the nodes and interactions with neighbors are linked by edges as shown in Fig.~\ref{figure1}(a). Edge weights characterize the degree of intimacy between individuals on the network. Specifically, if $\omega_{ij}=0$, there is no connection between individual $i$ and $j$, and no interaction occurs between them. While if $\omega_{ij}>0$, individual $i$ and $j$ are connected and can interact with each other. Here, graph $G$ is undirected and self-loops are not allowed. The weighted degree of vertex $i$ is indicated by $\omega_i=\sum_{j\in G}{\omega_{ij}}$ and the total weighted degree value of all vertices is expressed as $W=\sum_{i\in G}{\omega_i}=\sum_{i,j\in G}{\omega_{ij}}$.

During the evolutionary process, each individual will choose between the optional opinions $A$ and $B$, which are marked as $1$ and $0$, respectively. In this way, a particular state of all individuals can be represented by a vector $\textbf{s}=(s_i)_{i\in G}$, where $s_i\in \{0,1\}$ denotes the opinion that individual $i$ chooses. When an individual chooses an opinion, we assume that he/she can obtain a basic score about the chosen opinion based on the information of public knowledge, which is set as $\delta_A$ for opinion $A$ and $\delta_B$ for opinion $B$. Here, the basic score
is derived from the information of public knowledge. Alternatively, we use the basic score value to represent the preference strength on each opinion. Besides, we assume that each individual has the identical basic score on the same opinion. In addition, each individual with the chosen opinion can interact with his/her neighbors via pairwise interactions~\cite{Szabo1998PRE}. Such interactions can induce some feedback scores for all individuals. To be specific, we assume that when an individual with opinion $A(B)$ interacts with a neighbor who shares the same opinion $A(B)$, he/she can receive a positive score $a$ ($d$) due to the positive feedback effect. Otherwise, when an individual with opinion $A(B)$ interacts with a neighbor having the opposite opinion $B(A)$, he/she gets a negative score $b~(c)$. Here, we use these feedback score values to characterize the opinion interactions between two connected individuals. The score matrix for the pairwise interactions~\cite{Tadelis2013,Weibull1997,Hofbauer1998,Sigmund1999CB} between two individuals can be thus described as
\begin{equation}
\bordermatrix{%
      & A      & B    \cr
A     & a      & b    \cr
B     & c      & d
},\label{payoff}
\end{equation}
where the $a$ and $d$ matrix elements are positive representing positive feedback score values, while elements $b$ and $c$ are negative because of adverse response. Accordingly, each individual interacts with the neighbors and can obtain an accumulative score as the evaluation result on local opinion interactions with all the neighbors.

After interacting with all nearest neighbors, each individual $i$ receives a total score $f_i(\textbf{s})$, which represents the overall evaluation of the chosen opinion of individual $i$, taking into account public information and feedback from interactions with neighbors. Hence the total score $f_i(\textbf{s})$ is expressed as the sum of the basic score and the accumulative scores from opinion interactions, given as
\begin{equation}
f_i(\textbf{s})=s_i\delta_A+(1-s_i)\delta_B+P_i(\textbf{s}),
\end{equation}
where $P_i(\textbf{s})$ represents the edge-weighted accumulated score value of individual $i$ in state $\textbf{s}$. Here, $P_i(\textbf{s})$ means that the score values obtained from neighbors are multiplied by the corresponding edge weights and then summed, as illustrated in Fig.~\ref{figure1}(b), which depicts opinion interactions of individual $i$ with neighbors. Accordingly, $P_i(\textbf{s})$ can be written as
\begin{align}
P_i(\textbf{s})&=W\pi_i\left(as_is^{(1)}_i+bs_i\left(1-s_i^{(1)}\right)\right)\notag\\
&+W\pi_i\left(c\left(1-s_i\right)s^{(1)}_i+d(1-s_i)\left(1-s_i^{(1)}\right)\right)\notag\\
&=W\pi_i\left(\left(a-b-c+d\right)s_is^{(1)}_i+(b-d)s_i\right)\notag\\
&+W\pi_i\left(\left(c-d\right)s^{(1)}_i+d\right),
\end{align}
where $s^{(1)}_i=\Sigma^N_{i=1}p_{ij}s_j$ describes the expected type of nearest neighbors of individual $i$, $p_{ij}=\omega_{ij}/\omega_i$ represents the probability of the step from $i$ to $j$ for a random walk on $G$, and $\pi_i=\omega_i/W$ can be understood as the stationary probability of $i$ in the stationary distribution of random walks on $G$~\cite{Allen17Nature}.

Furthermore, we have
\begin{align}
f_i(\textbf{s})&=W\pi_i\left(\left(a-b-c+d\right)s_is^{(1)}_i+(b-d)s_i\right)\notag\\
&+W\pi_i\left(\left(c-d\right)s^{(1)}_i+d\right)+s_i\delta_A+(1-s_i)\delta_B.
\end{align}

Subsequently, each individual $i$ can update his/her opinion choice based on the total score information. By following  previous studies about binary opinions~\cite{Yang2016EPL,Zhou2018EPL}, the microscopic updating procedure of opinions on networks is governed by the ``pairwise comparison" updating in our study~\cite{Szabo1998PRE,Ohtsuki2006Nature,Ohtsuki2006JTB,Ohtsuki2008JTB}. To be specific, individual $i$ first selects a neighbor as the model player proportionally to the edge weights and accordingly neighbor $j$ is chosen with probability $p_{ij}$. Then individual $i$ imitates the opinion of the chosen neighbor $j$ with probability $F(s_i, s_j)$, given as
\begin{equation}
F(s_i,s_j)=\frac{1}{1+\exp[-\beta(f_j(\textbf{s})-f_i(\textbf{s}))]},
\end{equation}
where $\beta$ denotes the intensity of selection. For $\beta\rightarrow0$, the selection is weak~\cite{Konno2011JTB,Sample2017JMB}, which means that the score value is only a small perturbation to the neutral drift that is a baseline at $\beta=0$~\cite{Chen2013AAP}. In contrast, for $\beta\rightarrow +\infty$ limit the selection is strong in the sense that individual $i$ will deterministically imitate the opinion of his/her neighbors with higher score values~\cite{Santos2006PNAS,Santos2008Nature,Santos2012JTB,Fudenberg2006TPB}. Under the pairwise comparison updating procedure in our model, the opinion with higher scores held by neighbors are more likely be imitated, but the reversed process may also happen with a low probability. Hence, this so-called weak selection assumption which we consider in this study is reasonable, because beside currently discussed score, the imitation randomness or error is also considered in this case, which is reported in the spreading process of realistic situations~\cite{Zhang2021IEEETIFS}.

\section{Results}
Based on the above description, we can consider the evolutionary process as a continuous-time {\it Markov chain} $(\textbf{S}(t))_{t\geq0}$ and call it as {\it evolutionary Markov chain}, in which state transitions occur via imitation events~\cite{Allen17Nature,Allen2014JMB}. We indicate the imitation event that individual $i$ randomly imitates the opinion of neighboring $j$ by $j\rightarrow i$. Accordingly, the $j\rightarrow i$ event occurs with probability $R[j\rightarrow i]$, given as
\begin{equation}
{\rm R}[j\rightarrow i]=p_{ij}F(s_i,s_j).
\end{equation}

Hence the probability that individual $i$ keeps the original opinion is given as
\begin{equation}
{\rm R}[i\rightarrow i]=1-\sum_{j}{p_{ij}F(s_i,s_j)}.
\end{equation}

The {\it evolutionary Markov chain} in our model will be absorbed in one of the two states: all-$A$ and all-$B$, which represent the fixation of $A$ and $B$, respectively. Notably, the system spends only a short intermediate time in mixed states~\cite{Allen2014JMB}. Here, we focus on the fixation probability $\rho_A$ of opinion $A$ under weak selection~\cite{Allen17Nature}, which represents the probability that the population state eventually reaches the absorbing state of all-$A$ from an initial state $\textbf{s}_0$ in which there are only one individual with opinion $A$ and $N-1$ individuals with opinion $B$. We can obtain the mathematical expression of fixation probability $\rho_A$ (for detailed derivations see Appendix), given as
\begin{equation}
 \rho_A=\frac{1}{N}+\beta\langle D'(\textbf{s})\rangle_\textbf{u}^\circ+o(\beta^2),
\end{equation}
where $D'(\textbf{s})$ is the first-order term of the Taylor-expansion of $D(\textbf{s})$ at $\beta=0$ and $D(\textbf{s})$ represents the instantaneous rate of change in the degree-weighted frequency of opinion $A$ from state $\textbf{s}$. Here $\langle D'(\textbf{s})\rangle_{\textbf{u}}^\circ$ is expressed as
\begin{align}
&\left\langle D'(\textbf{s})\right\rangle_\textbf{u}^\circ \notag\\
&=\frac{W}{2}\left(b-d\right)\sum_{i,j}{\pi_i^2p_{ij}\left\langle s_i^2-s_is_j\right\rangle_\textbf{u}^\circ}\notag\\
&+\frac{W}{2}\left(a-b-c+d\right)\sum_{i,j,k}{\pi_i^2p_{ij}p_{ik}\left\langle s_i^2s_k-s_is_js_k\right\rangle_\textbf{u}^\circ}\notag\\
&+\frac{W}{2}\left(c-d\right)\sum_{i,j,k}{\pi_i^2p_{ij}p_{ik}\left\langle s_is_k-s_js_k\right\rangle_\textbf{u}^\circ}\notag\\
&+\frac{Wd}{2}\sum_{i,j}{\pi_i^2p_{ij}\left\langle s_i-s_j\right\rangle_\textbf{u}^\circ}\notag\\
&+\frac{\delta_A-\delta_B}{2}\sum_{i,j}{\pi_ip_{ij}\left\langle s_i^2-s_is_j\right\rangle_\textbf{u}^\circ}\label{state function},
\end{align}
where
%\begin{gather*}
\begin{align}
\langle s_i\rangle_\textbf{u}^\circ&=\int_0^{\infty}{E_\textbf{u}^\circ[S_i(t)]\mathrm{d}t},\\
\langle s_is_j\rangle_\textbf{u}^\circ&=\int_0^{\infty}{E_\textbf{u}^\circ[S_i(t)S_j(t)]\mathrm{d}t},
\end{align}
and
\begin{align}
\langle s_is_js_k\rangle_\textbf{u}^\circ&=\int_0^{\infty}{E_\textbf{u}^\circ[S_i(t)S_j(t)S_k(t)]\mathrm{d}t}.
%\end{gather*}
\end{align}
For calculating the quantities in Eq.~\eqref{state function}, we use the concept of coalescence times which will be introduced in the next subsection.

\subsection{Coalescing random walks}
In order to calculate $\langle s_i\rangle_\textbf{u}^\circ$, $\langle s_is_j\rangle_\textbf{u}^\circ$, and $\langle s_is_js_k\rangle_\textbf{u}^\circ$, we consider coalescing random walks on graphs~\cite{Kingman1982SPA,Wakeley2009,Allen17Nature,Cox1989AP}. Coalescing random walks on graph $G$ are defined as a collection of random walks, which is a process corresponding to tracing backwards ancestors. Specifically, if individual $i$ imitates the opinion of neighboring $j$ at a certain step during the evolutionary process, then individual $j$ is called the ``ancestor" of individual $i$ at that corresponding step.

During the evolutionary process in our model, individual $i$ imitates the opinion of individual $j$ under neutral drift ($\beta=0$) with probability $\tilde{p}_{ij}=p_{ij}\frac{1}{1+exp[-\beta(f_j(\textbf{s})-f_i(\textbf{s}))]}=\frac{1}{2}p_{ij}$ when $j\neq i$, hence individual $i$ keeps the original opinion with probability $\tilde{p}_{ii}=1-\sum _{j}{\tilde{p}_{ij}}=1-\sum_j{\frac{1}{2}p_{ij}}=\frac{1}{2}$. Therefore a new graph $G'$ can be generated based on graph $G$, on which self-loops are introduced due to the possible opinion keeping by some individuals. We can then note that for a random walk on $G'$, a step from $i$ to $j$ is taken with probability $\tilde{p}_{ij}=\frac{1}{2}p_{ij}$ for $ j\neq i$, and a step from $i$ to $i$ is taken with probability $\tilde{p}_{ii}=\frac{1}{2}$. Hence, the case under neutral drift in our model is dual to the continuous-time coalescing random walks on $G'$ as depicted in Fig.~\ref{figure1}(c) and (d). Furthermore, we need to consider one-, two-, and three-dimensional coalescing random walks on $G'$ to calculate the values of $\langle s_i\rangle_\textbf{u}^\circ$, $\langle s_is_j\rangle_\textbf{u}^\circ$, and $\langle s_is_js_k\rangle_\textbf{u}^\circ$, respectively.

\subsection{One-dimensional coalescing random walks}
For one-dimensional coalescing random walks on $G'$, we assume that there is a walker $X(t)$ where ${t\ge0}$, which occupies
an arbitrary vertex $i$ at the initial time, i.e., $X(0)=i$. At each time step, $X(t)$ takes a random step on $G'$. Each step is taken with the probability corresponding to the rate at which the imitation event occurs in the continuous-time evolutionary process.

As stated above, the case under neutral drift in our model is dual to the continuous-time coalescing random walks on $G'$. We thus can use the following equation to illustrate the duality relationship: for any initial state $\textbf{s}_0$ and any two types of opinions $m,n\in{\{1,0\}}$,
\begin{equation}
P_{\textbf{s}_0}^\circ[S_i(t)=m]=P_i^{CRW}[(\textbf{s}_0)_{X(t)}=m] \label{one duality},
\end{equation}
where $P_{\textbf{s}_0}^\circ[]$ denotes the probability value in the neutral evolutionary process started from state $s_0$, and $P_i^{CRW}[ ]$ represents the probability value in one-dimensional coalescing random walks started from $i$. Eq.~\eqref{one duality} means that in the neutral case of evolutionary process, the probability of individual $i$ with type $m$ is the same to the probability that the walker $X(t)$ steps to the ancestor of individual $i$ at time $t$ in the coalescing random walks started from $i$. In other words, the type of individual $i$ at time $t$ is identical to the type of his/her ancestor at the initial state.

For a special initial state $\textbf{s}_0$ satisfying $(\textbf{s}_0)_l=1$ and $(\textbf{s}_0)_k=0$ for all $k\neq l$, according to Eq.~\eqref{one duality} we have
\begin{align}
E_{\textbf{s}_0}^\circ[S_i(t)]&=P_{\textbf{s}_0}^\circ[S_i(t)=1]\notag\\
&=P_i^{CRW}[(\textbf{s}_0)_{X(t)}=1]
=P_i^{CRW}[X(t)=l] \label{initial one duality},
\end{align}
where $E_{\textbf{s}_0}^\circ[]$ denotes the expectation in the neutral evolutionary process started from state $s_0$.

We now define a probability distribution $\textbf{u}$ over all states $\textbf{s}$, assigning probability $\frac{1}{N}$ to the special states in which there is only one vertex with opinion $A$, and probability zero to other states. When the initial state is sampled from the probability distribution $\textbf{u}$, according to Eq.~\eqref{initial one duality} we have
\begin{equation}
E_\textbf{u}^\circ[S_i(t)]=\frac{1}{N}\sum _lP_i^{CRW}[X(t)=l]=\frac{1}{N}.
\end{equation}

Furthermore, we obtain $\langle s_i\rangle_\textbf{u}^\circ$ as
\begin{equation}
\langle s_i\rangle_\textbf{u}^\circ=\int_0^{\infty}{E_\textbf{u}^\circ[S_i(t)]\mathrm{d}t}=\int_0^{\infty}{\frac{1}{N}\mathrm{d}t} \label{one coalescence time}.
\end{equation}

\subsection{Two-dimensional coalescing random walks}
For two-dimensional coalescing random walks on $G'$, we consider continuous- and discrete-time versions, respectively~\cite{Allen17Nature}. In the continuous-time coalescing random walks, we assume that there are two walkers $(X(t), Y(t))_{t\geq0}$, which occupy arbitrarily chosen $i$ and $j$ vertices at the initial time, i.e., $X(0)=i$ and $Y(0)=j$. At each time step, they walk independently until they meet for the first time (coalescence). The first meeting time in two-dimensional coalescing random walks on $G'$ is denoted by $T_{coal}^{(2)}$. After this time, they walk together, that is, $X(t)=Y(t)$ for all $t>T_{coal}^{(2)}$. We define $P_{(i,j)}^{CRW}[ ]$ and $E_{(i,j)}^{CRW}[ ]$ to respectively represent the probabilities and expectations in continuous-time coalescing random walks started from $i$ and $j$.

In the discrete-time coalescing random walks, we also assume that there are two walkers $(X(t), Y(t))_{t=0}^\infty$, which occupy
randomly chosen $i$ and $j$ vertices at the initial time, i.e., $X(0)=i$ and $Y(0)=j$. Unlike continuous-time coalescing random walks, at each time step if $X(t)\neq Y(t)$, then one of $X(t)$ and $Y(t)$ will be randomly chosen to make a random step until their time of coalescing. If $X(t)=Y(t)$, they will take the same step at the next time step. We define $\tilde{P}_{(i,j)}^{CRW}[ ]$ and $\tilde{E}_{(i,j)}^{CRW}[ ]$ to respectively represent the probabilities and expectations in discrete-time and two-dimensional coalescing random walks on $G'$ started from $i$ and $j$.

Obviously, it can be seen that in the continuous-time version two steps are taken per unit time, while in the discrete-time version one step is taken every time step. In the discrete-time coalescing random walk started from $i$ and $j$, the expected coalescence time is defined by $\tau_{ij}=\tilde{E}_{(i,j)}^{CRW}[T_{coal}^{(2)}]$. Due to the different numbers of steps per unit time between the two versions, we obtain the expected coalescence time in continuous-time coalescing random walks $E_{(i,j)}^{CRW}[T_{coal}^{(2)}]=\tau_{ij}/2$.

Due to the fact that the neutral drift case of our model is dual to the continuous-time coalescing random walks on $G'$, for any initial state $\textbf{s}_0$ and any two types of opinions $m,n\in{\{1,0\}}$ we have
\begin{align}
&\quad P_{\textbf{s}_0}^\circ[S_i(t)=m, S_j(t)=n]\notag\\
&=P_{(i,j)}^{CRW}[(\textbf{s}_0)_{X(t)}=m, (\textbf{s}_0)_{Y(t)}=n] \label{two duality},
\end{align}
which indicates that the types of individual $i$ and $j$ at time $t$ in the evolutionary process under neutral drift are the same to their corresponding ancestors at the initial state.

Furthermore, we consider a special initial state $\textbf{s}_0$ satisfying $(\textbf{s}_0)_l=1$ and $(\textbf{s}_0)_k=0$ for all $k\neq l$. According to Eq.~\eqref{two duality}, we have
\begin{equation}
\begin{split}
E_{\textbf{s}_0}^\circ[S_i(t)S_j(t)]&=P_{\textbf{s}_0}^\circ[S_i(t)=1, S_j(t)=1]\\
&=P_{(i,j)}^{CRW}[X(t)=Y(t)=l]\\
&=P_{(i,j)}^{CRW}[T_{coal}^{(2)}<t,X(t)=l] \label{initial two duality}.
\end{split}
\end{equation}
When the initial state is sampled from the probability distribution $\textbf{u}$, we have
\begin{equation}
\begin{split}
E_\textbf{u}^\circ[S_i(t)S_j(t)]&=\frac{1}{N}\sum _lP_{(i,j)}^{CRW}[T_{coal}^{(2)}<t,X(t)=l]\\
&=\frac{1}{N}P_{(i,j)}^{CRW}[T_{coal}^{(2)}<t],
\end{split}
\end{equation}
where the second equality can be derived by the law of total probability.

Based on the calculation of $E_\textbf{u}^\circ[S_i(t)S_j(t)]$, we can have
\begin{equation}
\begin{split}
\left\langle\frac{1}{N}-s_is_j\right\rangle_\textbf{u}^\circ&=\int_0^{\infty}{\left(\frac{1}{N}-E_\textbf{u}^\circ[S_i(t)S_j(t)]\right)\mathrm{d}t}\\
&=\frac{1}{N}\int_0^{\infty}{\left(1-P_{(i,j)}^{CRW}[T_{coal}^{(2)}<t]\right)\mathrm{d}t}\\
&=\frac{1}{N}E_{(i,j)}^{CRW}[T_{coal}^{(2)}]\\
&=\frac{\tau_{ij}}{2N},
\end{split}
\end{equation}
where $\tau_{ij}/2$ denotes the expected coalescence time of continuous-time coalescing random walks.

Accordingly, we obtain the expression of $\langle s_is_j\rangle_\textbf{u}^\circ$ as
\begin{equation}
\langle s_is_j\rangle_\textbf{u}^\circ=\int_0^{\infty}{\frac{1}{N}\mathrm{d}t}-\frac{\tau_{ij}}{2N} \label{two}.
\end{equation}
The expected coalescence time $\tau_{ij}$ of two-dimensional and discrete-time coalescing random walks started from $i$ and $j$ satisfies the recurrence relation ~\cite{Allen17Nature}
\begin{equation}
\tau_{ij}=\left\{
\begin{array}{lcl}
0&&{i=j}\\
1+\frac{1}{2}\sum_{x\in G}{(\tilde{p}_{ix}\tau_{jx}+\tilde{p}_{jx}\tau_{ix}})&&{i\neq j},
\end{array}\right. \label{two coalescence time}
\end{equation}
where
Eq.~\eqref{two coalescence time} is a system of ${N\choose2}$ linear equations.

\subsection{Three-dimensional coalescing random walks}

Similar to the two-dimensional case, for three-dimensional coalescing random walks on $G'$ we also consider continuous- and discrete-time versions. In the continuous-time coalescing random walks, we assume that there are three walkers $(X(t), Y(t), Z(t))_{t\geq0}$, which occupy
randomly chosen $i$, $j$, and $k$ vertices at the initial time, i.e., $X(0)=i$, $Y(0)=j$, and $Z(0)=k$. At each time step, they walk independently until they meet for the first time. The first meeting time is represented by $T_{coal}^{(3)}$. After this time, they will walk together, i.e., $X(t)=Y(t)=Z(t)$ for $t>T_{coal}^{(3)}$. Particularly, if any two of them meet before $T_{coal}^{(3)}$, then the two walkers who have met will walk together afterwards. We define $P_{(i,j,k)}^{CRW}[ ]$ and $E_{(i,j,k)}^{CRW}[ ]$ to respectively denote the probabilities and expectations in the continuous-time and three-dimensional coalescing random walks on $G'$ started from $i$, $j$, and $k$.

For discrete-time coalescing random walks, we also assume that there are three walkers $(X(t), Y(t),Z(t))_{t=0}^\infty$, which occupy arbitrary three $i$, $j$, and $k$ vertices at the initial time, i.e., $X(0)=i$, $Y(0)=j$, and $Z(0)=k$. Unlike continuous-time coalescing random walks, at each time step if $X(t)$, $Y(t)$, and $Z(t)$ occupy different vertices, one of $X(t)$, $Y(t)$, and $Z(t)$ will be randomly chosen to make a random step until their time of coalescing. Particularly, if any two of them meet before coalescing, then these walkers stay together. Evidently, if $X(t)=Y(t)=Z(t)$, they all stay together in the following time. As previously, here we define $\tilde{P}_{(i,j,k)}^{CRW}[ ]$ and $\tilde{E}_{(i,j,k)}^{CRW}[ ]$ to respectively represent the probabilities and expectations in the discrete-time and three-dimensional coalescing random walks on $G'$ started from $i$, $j$, and $k$.

Obviously, it can be seen that in a continuous-time three-dimensional coalescing random walk three steps are taken per unit time, while in a discrete-time coalescing random walk one step is taken every time. In the discrete-time version, the expected coalescence time of three-dimensional coalescing random walks started from $i$, $j$, and $k$ is defined by $\tau_{ijk}=\tilde{E}_{(i,j,k)}^{CRW}[T_{coal}^{(3)}]$. Since the different numbers of steps per unit time between two versions, we obtain the expected coalescence time in continuous-time coalescing random walks
$E_{(i,j,k)}^{CRW}[T_{coal}^{(3)}]=\tau_{ijk}/3$.

Similarly, the three-dimensional coalescing random walks in the continuous-time version have the same relationship with the neutral drift case in our model, as found in two-dimensional coalescing random walks. For any initial state $\textbf{s}_0$ and any two types $m, n\in{\{1,0\}}$, we thus have
\begin{align}
&\quad P_{\textbf{s}_0}^\circ[S_i(t)=m, S_j(t)=n, S_k(t)=m]\notag\\
&=P_{(i,j,k)}^{CRW}[(\textbf{s}_0)_{X(t)}=m, (\textbf{s}_0)_{Y(t)}=n, (\textbf{s}_0)_{Z(t)}=m] \label{three duality},
\end{align}
which indicates that the types of individual $i$, $j$, and $k$ at time $t$ in the evolutionary process under neutral drift are the same to their corresponding ancestors at the initial state.

For a special initial state $\textbf{s}_0$ satisfying $(\textbf{s}_0)_l=1$ and $(\textbf{s}_0)_k=0$ for all $k\neq l$, we have
\begin{equation}
\begin{split}
E_{s_0}^\circ[S_i(t)S_j(t)S_k(t)]&=P_{(i,j,k)}^{CRW}[X(t)=Y(t)=Z(t)=l]\\
&=P_{(i,j,k)}^{CRW}[T_{coal}^{(3)}<t,X(t)=l] \label{initial three duality}.
\end{split}
\end{equation}
If the initial state is chosen from $\textbf{u}$, according to Eq. \eqref{initial three duality} we obtain
\begin{equation}
\begin{split}
E_\textbf{u}^\circ[S_i(t)S_j(t)S_k(t)]&=\frac{1}{N}\sum _lP_{(i,j,k)}^{CRW}[T_{coal}^{(3)}<t,X(t)=l]\\
&=\frac{1}{N}P_{(i,j,k)}^{CRW}[T_{coal}^{(3)}<t].
\end{split}
\end{equation}
Based on the calculation of $E_\textbf{u}^\circ[S_i(t)S_j(t)S_k(t)]$, we can have
\begin{equation}
\begin{split}
\left\langle\frac{1}{N}-s_is_js_k\right\rangle_\textbf{u}^\circ&=\int_0^{\infty}{\left(\frac{1}{N}-E_\textbf{u}^\circ[S_i(t)S_j(t)S_k(t)]\right)\mathrm{d}t}\\
&=\frac{1}{N}\int_0^{\infty}{\left(1-P_{(i,j,k)}^{CRW}[T_{coal}^{(3)}<t]\right)\mathrm{d}t}\\
&=\frac{1}{N}E_{(i,j,k)}^{CRW}[T_{coal}^{(3)}]\\
&=\frac{\tau_{ijk}}{3N}.
\end{split}
\end{equation}
Accordingly we obtain the mathematical expression of $\langle s_is_js_k\rangle_\textbf{u}^\circ$ as
\begin{equation}
\langle s_is_js_k\rangle_\textbf{u}^\circ=\int_0^{\infty}{\frac{1}{N}\mathrm{d}t}-\frac{\tau_{ijk}}{3N} \label{three}.
\end{equation}

The expected coalescence time $\tau_{ijk}$ of discrete-time and three-dimensional coalescing random walks started from $i,~j,~ k$ satisfies the recurrence relation
\begin{align}
&\tau_{ijk}=\notag\\
&\begin{cases}
0 &{i=j=k} \cr
1+\frac{1}{3}\sum_{x\in G}{(2\tilde{p}_{ix}\tau_{xxk}+\tilde{p}_{kx}\tau_{iix})} &{i=j\neq k} \cr
1+\frac{1}{3}\sum_{x\in G}{(\tilde{p}_{ix}\tau_{xjj}+2\tilde{p}_{jx}\tau_{ixx})} &{i\neq j=k} \cr
1+\frac{1}{3}\sum_{x\in G}{(\tilde{p}_{jx}\tau_{ixi}+2\tilde{p}_{kx}\tau_{xjx})} &{i=k\neq j} \cr
1+\frac{1}{3}\sum_{x\in G}{(\tilde{p}_{ix}\tau_{xjk}+\tilde{p}_{jx}\tau_{ixk}+\tilde{p}_{kx}\tau{_{ijx}})} &{i\neq j\neq k},
\end{cases}\ \label{three coalescence time}
\end{align}
where Eq.~\eqref{three coalescence time} is a system of ${N\choose3}$ linear equations.

\subsection{Condition for $\rho_A>\frac{1}{N}$}

Applying Eqs.~\eqref{state function},~\eqref{one coalescence time},~\eqref{two}, and~\eqref{three}, we obtain the fixation probability of opinion $A$ under weak selection as
\begin{equation}
\rho_A=\frac{1}{N}+\beta\langle D'(\textbf{s})\rangle_\textbf{u}^\circ+o(\beta^2),
\end{equation}
where
\begin{equation}
\begin{split}
&\left\langle D'(\textbf{s})\right\rangle_\textbf{u}^\circ\\
&=\frac{W}{2}\left(a-b-c+d\right)\sum_{i,j,k}\pi_i^2p_{ij}p_{ik}\frac{\tau_{ijk}-\tau_{iik}}{3N}\\
&+\frac{W}{2}\left(b-d\right)\sum_{i,j}\pi_i^2p_{ij}\frac{\tau_{ij}}{2N}\\
&+\frac{W}{2}\left(c-d\right)\sum_{i,j,k}\pi_i^2p_{ij}p_{ik}\frac{\tau_{jk}-\tau_{ik}}{2N}\\
&+\frac{\delta_A-\delta_B}{2}\sum_{i,j}\pi_ip_{ij}\frac{\tau_{ij}}{2N}.\label{D}
\end{split}
\end{equation}

For neutral selection, i.e., $\beta=0$, each individual imitates the opinion of the chosen neighbor with probability $0.5$ and accordingly the fixation probability of opinion A is $1/N$. Note that the case of neutral selection serves as a reference point~\cite{Allen17Nature}. In other words, if the fixation probability for $\beta\rightarrow 0$ is larger than $1/N$ for $\beta=0$, a single individual with opinion $A$ will invade and take over a population of individuals with opinion $B$. Hence opinion $A$ is favored to spread under weak selection, if and only if $\rho_A>\frac{1}{N}$, i.e.,
\begin{equation}
  \left\langle D'(\textbf{s})\right\rangle_\textbf{u}^\circ>0.
\end{equation}
From the above inequality, we can see that the successfully spreading condition is closely related to the basic scores of binary opinions, the feedback scores, and the structural parameters. Note that, in Eq.~\eqref{D}, $\pi_i=\frac{\omega_{i}}{W}$, where $\omega_{i}$ represents the weighted degree of vertex $i$ and $W$ is the sum of the weighted degrees of all vertices. In addition, $W$ can be expressed as $W=N\bar{w}$, where $\bar{w}$ denotes the average degree of the network. $p_{ij}=\frac{\omega_{ij}}{\omega_i}$, where $\omega_{ij}$ describes the edge weight. Therefore, the condition is closely related to the structural parameters including the edge weights, the weighted degrees of vertices, and the average degree of the network.

\subsection{Theoretical results for representative cases}

In this subsection, we consider three special but representative cases. For each case, we aim to derive the condition in which opinion $A$ is favored in the
weak selection limit.

\textbf{Case} \textbf{\uppercase\expandafter{\romannumeral1}}\textbf{:} We consider that individuals adjust their opinions only according to the basic evaluation score derived from the information of public knowledge. In other words, the positive and negative feedback scores from pairwise interactions are all zero, i.e., $a=b=c=d=0$. In this case, we find that opinion $A$ is favored under weak selection if and only if
\begin{equation}
\left\langle D'(\textbf{s})\right\rangle_\textbf{u}^\circ=\frac{\delta_A-\delta_B}{2}\sum_{i,j}\pi_ip_{ij}\frac{\tau_{ij}}{2N}>0.
\end{equation}
It can be seen that since $\sum_{i,j}\pi_ip_{ij}\frac{\tau_{ij}}{2N}>0$, thus whether opinion A can be favored to spread or not depends entirely on the difference of basic scores of A and B.  To be specific, opinion A is favored if and only if $\delta_A>\delta_B$.

\textbf{Case} \textbf{\uppercase\expandafter{\romannumeral2}}\textbf{:} We consider that $\delta_A=\delta_B$ and $b=c=0$. In this case, individuals adjust their opinions merely based on the positive feedback scores. Accordingly, we can obtain the mathematical condition in which opinion $A$ is favored, given as
\begin{align}
\frac{a}{d}&>(\frac{a}{d})^*\notag\\
&=\frac{3\sum_{i,j}\pi_i^2p_{ij}\tau_{ij}+3\sum_{i,j,k}\pi_i^2p_{ij}p_{ik}(\tau_{jk}-\tau_{ik})}{2\sum_{i,j,k}\pi_i^2p_{ij}p_{ik}(\tau_{ijk}-\tau_{iik})}-1\notag,\\
\label{ad}
\end{align}
which implies that when the ratio of the obtained positive scores of competing opinions exceeds the critical value $(\frac{a}{d})^*$, opinion A is favored to spread on the networks. Note that the critical value $(\frac{a}{d})^*$ depends on the structural parameters including the edge weights and weighted degrees of vertices.

In particular, for unweighed networks we have $\omega_{ij}\in \{0, 1\}$ and $\pi_i=\frac{w_i}{N\bar{w}}$, where $w_i$ denotes the degree of vertex $i$. Accordingly, $p_{ij}=\frac{1}{w_i}$ if $w_{ij}=1$, otherwise $p_{ij}=0$. In this case, Eq.~\eqref{ad} can be simplified to
\begin{align}
\frac{a}{d}&>\frac{3\sum_{i,j} w_i w_{ij}\tau_{ij}}{2\sum_{i,j,k}(\tau_{ijk}-\tau_{iik})w_{ij}w_{ik}}\notag\\
&+\frac{3\sum_{i,j,k}(\tau_{jk}-\tau_{ik})w_{ij}w_{ik}}{2\sum_{i,j,k}(\tau_{ijk}-\tau_{iik})w_{ij}w_{ik}}-1.
\end{align}
According to the above inequality, on unweighed networks we can conclude that the condition of successful spreading of opinion $A$ is related to the degrees of vertices in the networks.

\begin{figure*}[!t]
\begin{center}
\includegraphics[width=6.74in]{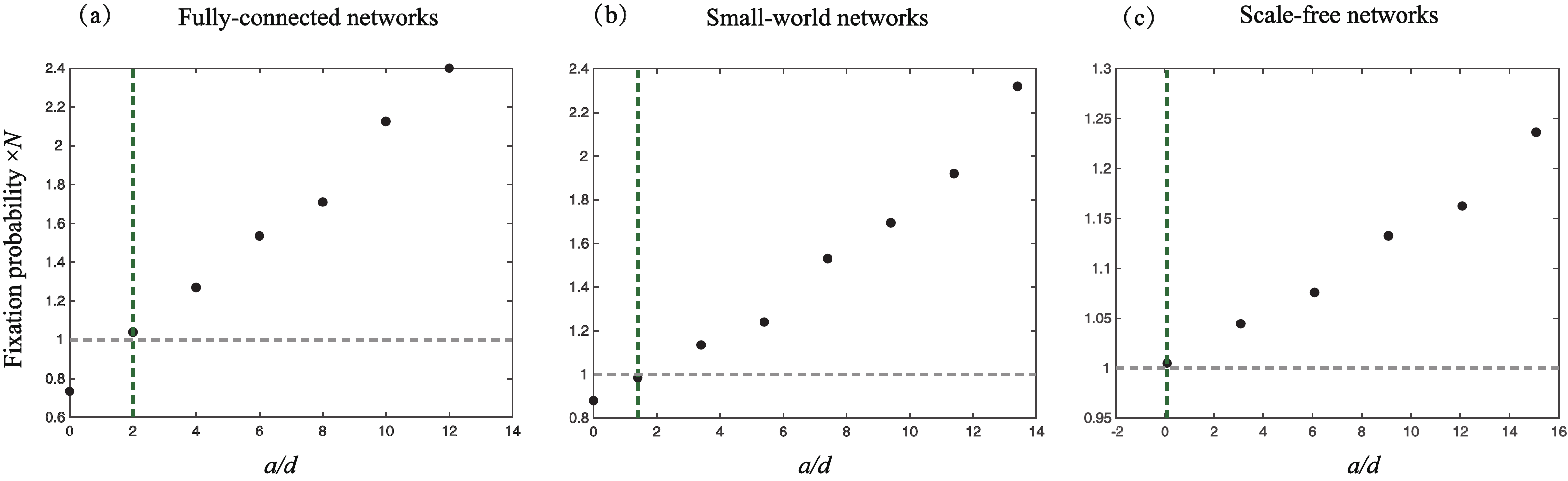}
\caption{The value of fixation probability $\rho_A$ multiplied by the system size $N$ as a function of the ratio $a/d$ for three representative graphs, including complete graph in panel (a), small-world networks with the initial number of neighbors $k=8$ and edge creation probability $p=0.4$ in panel (b), and scale-free networks with initial number of nodes $m_0=3$ and linking number $m=3$ in panel (c). The network size is all set to $N=50$ and all edge weight values are set to $1$ or $0$. Simulation results are represented by black solid circles and all green vertical dashed lines indicate the critical value $(a/d)^*$ obtained from Eq.~\eqref{ad}. When the $a/d$ value is larger than $(a/d)^*$, opinion $A$ is favored. Parameters are:
$\delta_A=\delta_B$ and $b=c=0$.}  \label{figure2}
\end{center}
\end{figure*}

\textbf{Case} \textbf{\uppercase\expandafter{\romannumeral3}}\textbf{:} We also consider the case where negative feedbacks are considered and set that $\delta_A=\delta_B$ and $a=d=0$. In this case, individuals adjust their opinions only based on the negative feedback scores. Accordingly, we can obtain the mathematical condition in which opinion $A$ is favored to spread, given as
\begin{align}
&\frac{b}{c}>(\frac{b}{c})^*\notag\\
&=\frac{2\sum_{i,j,k}\pi_i^2p_{ij}p_{ik}(\tau_{ijk}-\tau_{iik})-3\sum_{i,j,k}\pi_i^2p_{ij}p_{ik}(\tau_{jk}-\tau_{ik})}{3\sum_{i,j}\pi_i^2p_{ij}\tau_{ij}-2\sum_{i,j,k}\pi_i^2p_{ij}p_{ik}(\tau_{ijk}-\tau_{iik})}\notag.\\
\label{bc}
\end{align}
which implies that when the ratio of the obtained negative scores of competing opinions exceeds the critical value $(\frac{b}{c})^*$, opinion $A$ is favored to spread on the whole network. Note that the critical value $(\frac{b}{c})^*$ depends on the edge weights and weighted degrees of vertices.

In particular, for unweighed networks, Eq.~\eqref{bc} can be simplified to
\begin{align}
\frac{b}{c}&>\frac{2\sum_{i,j}(\tau_{ijk}-\tau_{iik})w_{ij}w_{ik}}{3\sum_{i,j} w_i\tau_{ij}w_{ij}-2\sum_{i,j,k}(\tau_{ijk}-\tau_{iik})w_{ij}w_{ik}}\notag\\
&-\frac{3\sum_{i,j,k}(\tau_{jk}-\tau_{ik})w_{ij}w_{ik}}{3\sum_{i,j} w_i\tau_{ij}w_{ij}-2\sum_{i,j,k}(\tau_{ijk}-\tau_{iik})w_{ij}w_{ik}}.
\end{align}
According to the above inequality, we can see that on unweighed networks, the condition of successful spreading of opinion $A$ is related to the degree values of vertices in the network.

\subsection{Simulation results }

In this subsection, to verify our theoretical results derived from Eq.~\eqref{ad}, we carry out Monte Carlo simulations on three representative
network structures including fully-connected, NW small-world~\cite{Newman1999PLA}, and BA scale-free networks~\cite{Barabasi1999Science}. All simulation results are obtained by performing the pairwise comparison updating rule. To reach the expected accuracy, we average the results over $10^4$ independent runs for a specific structure. In each simulation run, all individuals initially choose opinion $B$, except a randomly chosen individual who chooses opinion $A$. The fixation probability is approximated as the fraction of Monte Carlo simulation runs which eventuate in all-$A$ state. In Fig.~\ref{figure2}, we show the fixation probability multiplied by the system size $N$ as a function of the ratio $a/d$ for the three representative graphs. We find that the fixation probability of $A$ is close to $1/N$ to the greatest extent when $a/d$ reaches the theoretical critical value $(a/d)^*$ indicated by the green dashed vertical line. When the ratio exceeds $(a/d)^*$, the fixation probability is larger than $1/N$, which means that $A$ is favored to spread under weak selection. From Fig.~\ref{figure2}, we can see that all simulation results are in good agreement with our numerical results obtained from Eq.~\eqref{ad}.

\section{Conclusion}

Our basic motivation in this work is to provide a theoretical framework for a model of opinion dynamics where the microscopic rule of opinion update is based on a score value which is rooted from public knowledge about competing opinions and also depends on the interactions with neighbors. According to the obtained total score, each agent in the population can adjust individual opinion during the evolutionary process. Importantly,
we have studied the opinion dynamics on any network structures in the framework of our model. By means of theoretical analysis, we have derived
a mathematical expression of fixation probability of opinion $A$ and further obtained the condition in which opinion $A$ is favored under weak selection. We find that whether opinion $A$ is favored or not depends sensitively on the score parameters and the structural parameters including the edge weights, the weighted degrees of vertices, and the average degree of the network. In particular, when individuals adjust their opinions based only on the basic score, the diffusion of opinion $A$ depends on the difference of basic scores of opinion $A$ and $B$. In addition, we consider the special case in which the negative (positive) feedback effects of strategic interactions of opinions are ignored and find that there exists a critical value of the ratio $a/d$ ($b/c$) feedback parameters, above which opinion $A$ is favored under weak selection; interestingly, the value of this critical ratio is related to the edge weights and weighted degrees of vertices in the network and on unweighted networks it depends on the degree values of nodes. To complete our study, we also carry out computer simulations where we use the above specified interaction graphs. These simulation data are in good agreement with our theoretical predictions and confirm the robustness of our findings. We hope that our work will contribute to a deeper understanding of different spreading process on complex networks \cite{Tucker2018RSO,Amaral2020PRE,Arruda2018PR,Gimenez2021EPJB,Auer15SR}.

Recently, game-theoretical approaches have been introduced to study opinion dynamics on social networks~\cite{Wu2020CCC,Ding2009IJMPC,Ding2010PASMA,Di2007IJMPC,Yang2016EPL,Zhou2018EPL,Zino2020CDC}. It has been proven that the voter model on the evolving network is equivalent to the Moran process on the complete network in fixation probability and can be captured by the replicator equation in evolutionary game theory~\cite{Wu2020CCC}. These results demonstrate that game-theoretical approaches can pave the way for some challenging problems in opinion dynamics and can provide some game-theoretical insights into opinion formation~\cite{Wu2020CCC}. Similarly, by means of some elements of the technical approach used for evolutionary games on complex networks~\cite{Allen17Nature}, in this work we focus on the process of opinion propagation with strategic interactions. We theoretically obtain the condition in which one opinion can successfully invade a population and show that this result can be applied in different types of structured networks. We believe that our work can also shed some light on how opinion formation happens from a game-theoretical perspective.

Indeed, the basic scores derived from the public knowledge in our model can be regarded as a kind of baseline fitness  for opinion $A$ and $B$, hence our work can be a re-interpretation of previous works in evolutionary games from this viewpoint~\cite{Ohtsuki2006Nature,Allen17Nature,Hauser2014JTB}. However, it is worthy emphasizing that our work focuses on exploring opinion propagation with strategic interactions on social networks by using the game-theoretical approach. Importantly, the components of total score are also based on sociological considerations, which allow us to extend the concept originally used in related works of evolutionary games. To our best knowledge, models of opinion dynamics have never used the information combined from local and global sources, hence our work could potentially open a new research path in this field.

In this study, we mainly focus on the fixation probability in the initial condition where an individual at a vertex is chosen randomly to have opinion $A$ in the sea of individuals with opinion $B$. Indeed, we can investigate the fixation probability in other initial conditions. However, compared with other initial conditions, this initial condition we considered provides a harsher environment for the formation of opinion A. Accordingly, in this initial condition we can quantify how an opinion evolves and spreads in the sea of another different opinion through a bottom-up approach in an evolutionary process. Furthermore, using the quantity of the fixation probability in this condition, we can intuitively understand the spreading process in some realistic situations. For example, we can understand how a rumor grows out of nothing and finally spreads in a population and quantify how likely this outcome happens in game-based interactions. We can also understand how a virus spreads among a healthy population and quantify how likely this outcome occurs in game-based interactions.

In addition, we do not consider the issue of fixation time. Indeed, beside the fixation probability, the fixation time depicting the average time until the fixation occurs is another important quantity for studying evolutionary dynamics of binary opinions on social networks. In the framework of evolutionary games on graphs, some previous work have studied the fixation time of competing strategies on graphs~\cite{Hindersin2014JRSI,Askari2015PRE,Farhang2017PLOSCB,Hathcock2019PRE,Hindersin2019CB,Tkadlec2019CB,Sui2015PRE,Xiao2019PCB}.
For a future study it could be a promising extension to investigate the fixation time based on the theoretical approaches from evolutionary games on graphs.

\section*{Appendix}
\setcounter{equation}{0}
\renewcommand{\theequation}{A\arabic{equation}}

In the following, we mainly carry out the derivation of the fixation probability of opinion $A$. In order to describe the distribution of opinion $A$ in the population in state $\textbf{s}$, we define the degree-weighted frequency of opinion $A$ as
\begin{equation}
  \hat{s}=\sum_{i\in G}{\pi_is_i}.
\end{equation}

The degree-weighted frequency of $A$ at time $T$ can be denoted by a random variable $\hat{S}(T)$, given as
\begin{equation}
\hat{S}(T)=\sum_{i\in G}{\pi_iS_i(T)}.\label{weighted frequency}
\end{equation}
The weighting $\pi_i$ in Eq.~\eqref{weighted frequency} can be regraded as the reproductive value of vertex $i$ in evolutionary game theory, which quantifies the fixation probability of opinion $A$ under neutral drift~\cite{McAvoy2021JMB,Maciejewski2014JTB,Lessard2007JMB,Tarnita2014AN}.

We consider the {\it evolutionary Markov chain} started from arbitrary initial state $\textbf{S}(0)=\textbf{s}_0\in {\{0,1\}^G}$. According to the Fundamental Theorem of Calculus~\cite{Allen17Nature}, the expectation of degree-weighted frequency $E_{\textbf{s}_0}[\hat{S}(T)]$ satisfies
\begin{equation}
E_{\textbf{s}_0}[\hat{S}(T)]=\hat s_0+\int_0^T{\frac{\mathrm{d}}{\mathrm{d}t}E_{\textbf{s}_0}[\hat{S}(t)]\mathrm{d}t}.
\end{equation}

As stated above, the {\it evolutionary Markov chain} will be absorbed in the fixation states, all-$A$ or all-$B$, for any given initial state. Thus, in the limit $T\rightarrow\infty$ the expectation of degree-weighted frequency $E_{s_0}[\hat{S}(T)]$ is equivalent to the fixation probability of type $A$, that is,
\begin{equation}
  \rho_{\textbf{s}_0}=\hat s_0+\int_0^\infty{\frac{\mathrm{d}}{\mathrm{d}t}E_{\textbf{s}_0}[\hat{S}(t)]\mathrm{d}t} \label{fixation1}.
\end{equation}

In the following part, we will focus on the calculation of $\rho_{\textbf{s}_0}$. The crucial part is the integrand in Eq.~\eqref{fixation1}. Therefore, a state function $D(\textbf{s})$ is defined in this section, which describes the expected instantaneous rate of change about degree-weighted frequency of opinion $A$ in state $\textbf{s}$, exactly corresponding to the differential in the integrand. $D(\textbf{s})$ satisfies
\begin{equation}
E[\hat{S}(t+\varepsilon)-\hat{S}(t)\mid{S(t)=\textbf{s}}]=D(\textbf{s})\varepsilon+o(\epsilon)\quad(\varepsilon\rightarrow{0^+}) \label{D(s)}.
\end{equation}
Substituting Eq.~\eqref{D(s)} into Eq.~\eqref{fixation1}, we obtain
\begin{equation}
 \rho_{\textbf{s}_0}=\hat s_0+\int_0^{\infty}{E_{\textbf{s}_0}[D(\textbf {S}(t))]\mathrm{d}t} \label{fixation 2}.
\end{equation}

Note that it is challenging to compute $\rho_{\textbf{s}_0}$ exactly for arbitrary $s_0$. Therefore we here concentrate on the effects of weak selection on the fixation probability, meaning that $\beta\rightarrow 0$. In order to derive the fixation probability $\rho_{\textbf{s}_0}$ under weak selection, i.e., the first order in $\beta$ as $\beta\rightarrow 0^+$, we write the Taylor series expansion of $D(\textbf{s})$ in $\beta$ when $\beta\rightarrow 0$ as

\begin{equation}
\begin{split}
D(\textbf{s})&=D^\circ(\textbf{s})+\beta D'(\textbf{s})+o(\beta^2)\\
&=\beta D'(\textbf{s})+o(\beta^2) \label{taylor},
\end{split}
\end{equation}
where $D^\circ(\textbf{s})$ denotes the value of $D(\textbf{s})$ under neutral drift ($\beta=0$). Here, the superscript $^\circ$ is used to denote the case of neutral drift. We will show that $D^\circ(\textbf{s})=0$ for all $\textbf{s}\in {\{0,1\}^G}$ under the pairwise comparison updating in the following part.

The expansion of integrand in Eq.~\eqref{fixation 2} can be written as
\begin{equation}
\setlength{\abovedisplayskip}{16pt}
\setlength{\belowdisplayskip}{16pt}
\begin{split}
E_{\textbf{s}_0}[D(\textbf {S}(t))]&=\sum_\textbf{s}{P_{\textbf{s}_0}[\textbf{S}(t)=\textbf{s}]D(\textbf{s})}\\
&=\beta\sum_\textbf{s}{P_{\textbf{s}_0}^\circ[\textbf{S}(t)=\textbf{s}]D'(\textbf{s})}+o(\beta^2)\\
&=\beta E_{\textbf{s}_0}^\circ[D'(\textbf {S}(t))]+o(\beta^2) \label{expectation}.
\end{split}
\end{equation}
Substituting Eq.~\eqref{expectation} into Eq.~\eqref{fixation 2}, we obtain the formula of fixation probability under weak selection as
\begin{equation}
\setlength{\abovedisplayskip}{16pt}
\setlength{\belowdisplayskip}{16pt}
\rho_{\textbf{s}_0}=\hat s_0+\beta\int_0^{\infty}{E_{\textbf{s}_0}^\circ[D'(\textbf {S}(t))]\mathrm{d}t}+o(\beta^2) \label{fixation 3}.
\end{equation}
For convenience, we define a new operator $\langle \rangle_{\textbf{s}_0}^\circ$ to abbreviate the expression of fixation probability. Given any initial state $\textbf{s}_0$, for any function $g(\textbf{s})$ of state $\textbf{s}$ we have
\begin{equation}
\setlength{\abovedisplayskip}{16pt}
\setlength{\belowdisplayskip}{16pt}
\langle g\rangle_{\textbf{s}_0}^\circ=\int_0^{\infty}{E_{\textbf{s}_0}^\circ[g(\textbf {S}(t))]\mathrm{d}t}.
\end{equation}
Hence Eq.~\eqref{fixation 3} is abbreviated as
\begin{equation}
\setlength{\abovedisplayskip}{16pt}
\setlength{\belowdisplayskip}{16pt}
\rho_{\textbf{s}_0}=\hat s_0+\beta\langle D'\rangle_{\textbf{s}_0}^\circ+o(\beta^2) \label{fixation 4}.
\end{equation}
In order to derive the formula of fixation probability $\rho_{\textbf{s}_0}$, we have to calculate the expected instantaneous rate of the change in degree-weighted frequency $D(\textbf{s})$ for state $\textbf{s}$ in the weak selection limit. If the imitation event that individual $i$ imitates the opinion of his/her neighbor $j$ occurs, the degree-weighted frequency $\hat{s}$ will be changed by $\pi_i(s_j-s_i)$. Therefore, the expected instantaneous rate of degree-weighted frequency change from state $\textbf{s}$ is given by

\begin{flalign} \label{derivation}
&D(\textbf{s})=&\notag\\
&\sum_i\pi_i\left(-s_i+\sum_j{p_{ij}F(s_i,s_j)s_j+\left(1-\sum_j{p_{ij}F(s_i,s_j)}\right)s_i}\right)&\notag\\
&=\sum_i\pi_i\left(\sum_j{p_{ij}F(s_i,s_j)s_j-\sum_j{p_{ij}F(s_i,s_j)}s_i}\right)& \notag\\
&=\sum_i\pi_i\left(\sum_j{p_{ij}F(s_i,s_j)(s_j-s_i)}\right)& \notag\\
&=\sum_i\pi_i\left(\sum_j{p_{ij}(s_j-s_i)}\right)\left(\frac{1}{2}-\frac{\beta}{4}(f_i\left(\textbf{s})-f_j(\textbf{s})\right)\right)+o(\beta^2)&\notag\\
&=\sum_i{\sum_j{\pi_ip_{ij}s_j\left(\frac{1}{2}-\frac{\beta}{4}\left(f_i(\textbf{s})-f_j(\textbf{s})\right)\right)}}&\notag\\
&-\sum_i{\sum_j{\pi_ip_{ij}s_i\left(\frac{1}{2}-\frac{\beta}{4}\left(f_i(\textbf{s})-f_j(\textbf{s})\right)\right)}}+o(\beta^2)&\notag\\
&=\frac{1}{2}(\sum_i{\sum_j{\pi_ip_{ij}s_j}}-\sum_i{\sum_j{\pi_ip_{ij}s_i}})&\notag\\
&+\frac{\beta}{4}\sum_i{\sum_j{\pi_ip_{ij}s_i\left(f_i(\textbf{s})-f_j(\textbf{s})\right)}}&\notag\\
&-\frac{\beta}{4}\sum_i{\sum_j{\pi_ip_{ij}s_j\left(f_i(\textbf{s})-f_j(\textbf{s})\right)}}+o(\beta^2)&\notag\\
&=\frac{\beta}{4}\sum_i{\sum_j{\pi_ip_{ij}s_i\left(f_i(\textbf{s})-f_j(\textbf{s})\right)}}&\notag\\
&+\frac{\beta}{4}\sum_i{\sum_j{\pi_jp_{ji}s_j\left(f_j(\textbf{s})-f_i(\textbf{s})\right)}}+o(\beta^2)&\notag\\
&=\frac{\beta}{2}\sum_i{\pi_is_i\left(f_i(\textbf{s})-f_i^{(1)}(\textbf{s})\right)}+o(\beta^2).
\end{flalign}

Eq.~\eqref{derivation} implies $D^\circ(\textbf{s})=0$ for all states $\textbf{s}$ and here the superscript $^\circ$ denotes the case of neutral drift $\beta=0$. It can be seen that $D(\textbf{s})$ is related to the total score of individual $i$ and that of his/her one-step neighbor.

According to Eq.~\eqref{taylor}, we have
\begin{equation}\label{one order}
D'(\textbf{s})=\frac{1}{2}\sum_i{\pi_is_i\left(f_i(\textbf{s})-f_i^{(1)}(\textbf{s})\right)}.
\end{equation}

Note that the expected total score of the one-step neighbor of individual $i$ is given by
\begin{equation}
\begin{split}
&f_i^{(1)}(\textbf{s})=\sum_j{p_{ij}f_j(\textbf{s})}\\
&=W\sum_j{p_{ij}\pi_j\left(\left(a-b-c+d\right)s_js^{(1)}_j+(b-d)s_j\right)}\\
&+W\sum_j{p_{ij}\pi_j\left(\left(c-d\right)s^{(1)}_j+d\right)}\\
&+\sum_jp_{ij}{s_j\delta_A}+\sum_j{p_{ij}(1-s_j)\delta_B} \label{one step neighbor}.
\end{split}
\end{equation}

By substituting Eq.~\eqref{one step neighbor} into Eq.~\eqref{one order} we have

\begin{align}
&D'(\textbf{s})=\frac{1}{2}\sum_i{\pi_is_i\left(f_i(\textbf{s})-f_i^{(1)}(\textbf{s})\right)}\notag\\
&=\frac{W}{2}\left(a-b-c+d\right)\left(\sum_i{\pi_i^2s_i^2s_i^{(1)}}-\sum_i{\sum_j{\pi_i\pi_jp_{ij}s_is_js_j^{(1)}}}\right)\notag\\
&+\frac{W}{2}(b-d)(\sum_i{\pi_i^2s_i^2}-\sum_i{\sum_j{\pi_i\pi_jp_{ij}s_is_j}})\notag\\
&+\frac{W}{2}\left(c-d\right)\left(\sum_i{\pi_i^2s_is_i^{(1)}}-\sum_i{\sum_j{\pi_i\pi_jp_{ij}s_is_j^{(1)}}}\right)\notag\\
&+\frac{Wd}{2}(\sum_i{\pi_i^2s_i}-\sum_i{\sum_j{\pi_i\pi_jp_{ij}s_i}})\notag\\
&+\frac{\delta_A-\delta_B}{2}(\sum_i{\pi_is_i^2}-\sum_i{\sum_j{\pi_ip_{ij}s_is_j}}).
\end{align}
Since random walks have the reversibility property for each $i,j\in G$ and $\pi_ip_{ij}=\pi_jp_{ji}$, we have
\begin{align}
&D'(\textbf{s})=\frac{1}{2}\sum_i{\pi_is_i\left(f_i(\textbf{s})-f_i^{(1)}(\textbf{s})\right)}\notag\\
&=\frac{W}{2}\left(a-b-c+d\right)\left(\sum_i{\pi_i^2s_i^2s_i^{(1)}}-\sum_i{\sum_j{\pi_i^2p_{ij}s_is_js_i^{(1)}}}\right)\notag\\
&+\frac{W}{2}(b-d)(\sum_i{\pi_i^2s_i^2}-\sum_i{\sum_j{\pi_i^2p_{ij}s_is_j}})\notag\\
&+\frac{W}{2}\left(c-d\right)\left(\sum_i{\pi_i^2s_is_i^{(1)}}-\sum_i{\sum_j{\pi_i^2p_{ij}s_js_i^{(1)}}}\right)\notag\\
&+\frac{Wd}{2}(\sum_i{\pi_i^2s_i}-\sum_i{\sum_j{\pi_i^2p_{ij}s_j}})\notag\\
&+\frac{\delta_A-\delta_B}{2}(\sum_i{\pi_is_i^2}-\sum_i{\sum_j{\pi_ip_{ij}s_is_j}}),
\end{align}
where $s_i^{(1)}=\sum_k{p_{ik}s_k}$ denotes the expected type of the neighbor of individual $i$. Hence

\begin{align}\label{rate}
&D'(\textbf{s})=\frac{1}{2}\sum_{i}{\pi_is_i\left(f_i(\textbf{s})-f_i^{(1)}(\textbf{s})\right)}\notag\\
&=\frac{W}{2}\left(a-b-c+d\right)\sum_{i,j,k}{\pi_i^2p_{ij}p_{ik}\left(s_i^2s_k-s_is_js_k\right)}\notag\\
&+\frac{W}{2}\left(b-d\right)\sum_{i,j}{\pi_i^2p_{ij}\left(s_i^2-s_is_j\right)}\notag\\
&+\frac{W}{2}\left(c-d\right)\sum_{i,j,k}{\pi_i^2p_{ij}p_{ik}\left(s_is_k-s_js_k\right)}\notag\\
&+\frac{Wd}{2}\sum_{i,j}{\pi_i^2p_{ij}\left(s_i-s_j\right)}\notag\\
&+\frac{\delta_A-\delta_B}{2}\sum_{i,j}{\pi_ip_{ij}\left(s_i^2-s_is_j\right)}.
\end{align}
Accordingly, the fixation probability of $A$ is given by
\begin{equation}
\rho_{\textbf{s}_0}=\hat s_0+\beta\langle D'\rangle_{\textbf{s}_0}^\circ+o(\beta^2).
\end{equation}

We have obtained $\rho_{\textbf{s}_0}$ for any initial state $\textbf{s}_0$. Here we concentrate on the initial state where there is only one individual choosing $A$ in the population. We mainly compute the fixation probability for such a special initial state $\textbf{s}_0$ in which $s_i=1$ and $s_j=0$ for all $j\neq i$. Let $\textbf{u}$ be the probability distribution over all states $\textbf{s}$, assigning probability $\frac{1}{N}$ to states in which there is only one vertex with opinion $A$, and probability zero to all other states. Hence, when the initial state $\textbf{s}_0$  of the \emph{evolutionary Markov chain} is sampled from $\textbf{u}$, the fixation probability of opinion $A$ is given as
\begin{equation}
 \rho_A=\frac{1}{N}+\beta\langle D'\rangle_\textbf{u}^\circ+o(\beta^2),
\end{equation}
where
\begin{align}
&\left\langle D'(\textbf{s})\right\rangle_\textbf{u}^\circ \notag\\
&=\frac{W}{2}\left(a-b-c+d\right)\sum_{i,j,k}{\pi_i^2p_{ij}p_{ik}\left\langle s_i^2s_k-s_is_js_k\right\rangle_\textbf{u}^\circ}\notag\\
&+\frac{W}{2}\left(b-d\right)\sum_{i,j}{\pi_i^2p_{ij}\left\langle s_i^2-s_is_j\right\rangle_\textbf{u}^\circ}\notag\\
&+\frac{W}{2}\left(c-d\right)\sum_{i,j,k}{\pi_i^2p_{ij}p_{ik}\left\langle s_is_k-s_js_k\right\rangle_\textbf{u}^\circ}\notag\\
&+\frac{Wd}{2}\sum_{i,j}{\pi_i^2p_{ij}\left\langle s_i-s_j\right\rangle_\textbf{u}^\circ}\notag\\
&+\frac{\delta_A-\delta_B}{2}\sum_{i,j}{\pi_ip_{ij}\left\langle s_i^2-s_is_j\right\rangle_\textbf{u}^\circ}.
\end{align}
Here
%\begin{gather*}
\begin{align}
\langle s_i\rangle_\textbf{u}^\circ&=\int_0^{\infty}{E_\textbf{u}^\circ[S_i(t)]\mathrm{d}t},\\
\langle s_is_j\rangle_\textbf{u}^\circ&=\int_0^{\infty}{E_\textbf{u}^\circ[S_i(t)S_j(t)]\mathrm{d}t},
\end{align}
and
\begin{align}
\langle s_is_js_k\rangle_\textbf{u}^\circ&=\int_0^{\infty}{E_\textbf{u}^\circ[S_i(t)S_j(t)S_k(t)]\mathrm{d}t},
%\end{gather*}
\end{align}
where the subscript $\textbf{u}$ is used to denote the expectation of a quantity, when the initial state of the \emph{evolutionary Markov chain} is sampled from $\textbf{u}$.

\section*{Acknowledgments}

This research was supported by the National Natural Science Foundation of China (Grant Nos. 61976048, 62036002, and 61773121) and the Fundamental Research Funds of the Central Universities of China.

\section*{Author Declarations}
\subsection*{Conflict of Interest}
The authors have no conflicts to disclose.

\section*{Data Availability}
The data that support the findings of this study are available from the corresponding author upon reasonable request.


\begin{thebibliography}{00}

\bibitem{Acemoglu2011DGA} D. Acemoglu and A. Ozdaglar, \emph{Dyn. Games Appl.} \textbf{1}, 3-49 (2011).

\bibitem{Yildiz2013ACMTEC} E. Yildiz, A. Ozdaglar, D. Acemoglu, et al., \emph{ACM Trans. Econ. Comput.} \textbf{1}, 1-30 (2013).

\bibitem{Proskurnikov2017ARC} A. V. Proskurnikov and R. Tempo, \emph{Annu. Rev. Control} \textbf{43}, 65-79 (2017).

\bibitem{Lin2018IEEE TAC} X. Lin, Q. Jiao, and L. Wang, \emph{IEEE Trans. Autom. Control} \textbf{64}, 3431-3438 (2018).

\bibitem{Proskurnikov2015IEEE TAC} A. V. Proskurnikov, A. S. Matveev, and M. Cao, \emph{IEEE Trans. Autom. Control} \textbf{61}, 1524-1536 (2015).

\bibitem{Ghaderi2014Automatica} J. Ghaderi and R. Srikant, \emph{Automatica} \textbf{50}, 3209-3215 (2014).

\bibitem{Battiston2020PR} F. Battiston, G. Cencetti, I. Iacopini, et al., \emph{Phys. Rep.} \textbf{874}, 1-92 (2020).

\bibitem{DeGroot1974JASA} M. H. DeGroot, \emph{J. Am. Stat. Assoc.} \textbf{69}, 118-121 (1974).

\bibitem{Friedkin1990JMS} N. E. Friedkin and E. C. Johnsen, \emph{J. Math. Sociol.} \textbf{15}, 193-206 (1990).

\bibitem{Castellano2009RMP} C. Castellano, S. Fortunato, and V. Loreto, \emph{Rev. Mod. Phys.} \textbf{81}, 591-646 (2009).

\bibitem{Weisbuch2004EPJB} G. Weisbuch, \emph{Eur. Phys. J. B} \textbf{38}, 339-343 (2004).

\bibitem{Holcombe1998PC} R. G. Holcombe, \emph{Public Choice} \textbf{61}, 115-125 (1989).

\bibitem{Galam2002EPJBMCS} S. Galam, \emph{Eur. Phys. J. B} \textbf{25}, 403-406 (2002).

\bibitem{Sznajd-Weron2000IJMPC} K. Sznajd-Weron and J. Sznajd, \emph{Int. J. Mod. Phys. C} \textbf{11}, 1157-1165 (2000).

\bibitem{Deffuant2000ACS} G. Deffuant, D. Neau, F. Amblard, and G. Weisbuch, \emph{Adv. Complex Syst.} \textbf{3}, 87-98 (2000).

\bibitem{Krause2002} R. Hegselmann and U. Krause, \emph{J. Artif. Soc. Soc. Simul.} \textbf{5}, 1-33 (2002).

\bibitem{Hegselmann2005CEF} R. Hegselmann and U. Krause, \emph{Comput. Econ.} \textbf{25}, 381-405 (2005).

\bibitem{Jiang2014IEEETSP} C. Jiang, Y. Chen, and K. J. R. Liu, \emph{IEEE Trans. Signal Proces.} \textbf{62}, 4573-4586 (2014).

\bibitem{Zhang2021IEEETIFS} H. Zhang, Y. Li, Y. Chen, and H. V. Zhao, \emph{IEEE Trans. Info. Fore. Secu.} \textbf{16}, 1203-1217 (2021).

\bibitem{Di2007IJMPC} A. Di Mare and V. Latora, \emph{Int. J. Mod. Phys. C} \textbf{18}, 1377-1395 (2007).

\bibitem{Ding2009IJMPC} F. Ding, Y. Liu, and Y. Li, \emph{Int. J. Mod. Phys. C} \textbf{20}, 479-490 (2009).

\bibitem{Ding2010PASMA} F. Ding, Y. Liu, B. Shen, and X. M. Si, \emph{Phys. A} \textbf{389}, 1745-1752 (2010).

\bibitem{Yang2016EPL} H.-X. Yang, \emph{EPL} \textbf{115}, 40007 (2016).

\bibitem{Zhou2018EPL} J. L. Zhou, C. W. Huang, and Q. L. Dai, \emph{EPL} \textbf{123}, 30004 (2018).

\bibitem{Zino2020CDC} L. Zino, M. Ye, and M. Cao, \emph{59th IEEE Conference on Decision and Control}, 1110-1115 (2020).

\bibitem{Wu2020CCC} B. Wu, J. Du, and L. Wang, \emph{Proceedings of the 39th Chinese Control Conference}, 6707-6714 (2020).

\bibitem{Traulsen2007JTB} A. Traulsen, J. M. Pacheco, and M. A. Nowak, \emph{J. Theor. Biol.} \textbf{246}, 522-529 (2007).

\bibitem{Rabin1993AER} M. Rabin, \emph{Am. Econ. Rev.} \textbf{83}, 1281-1302 (1993).

\bibitem{Cao2008PRE} L. Cao and X. Li, \emph{Phys. Rev. E} \textbf{77}, 016108 (2008).

\bibitem{Szolnoki2015IF} A. Szolnoki and M. Perc, \emph{J. R. Soc. Interface} \textbf{12}, 20141299 (2015).

\bibitem{Szolnoki2018NJP} A. Szolnoki and X. Chen, \emph{New J. Phys.} \textbf{20}, 093008 (2018).

\bibitem{Weibull1997} J. W. Weibull, Evolutionary game theory, (MIT Press, 1997).

\bibitem{Hofbauer1998} J. Hofbauer and K. Sigmund, Evolutionary games and population dynamics, (Cambridge University Press, 1998).

\bibitem{Sigmund1999CB} K. Sigmund and M. A. Nowak, \emph{Curr. Biol.} \textbf{9}, 503-505 (1999).

\bibitem{Tadelis2013} S. Tadelis, Game theory: an introduction, (Princeton University Press, 2013).

\bibitem{Peng2018ASC} H. G. Peng, X. K. Wang, T. L. Wang, and J. Q. Wang, \emph{Appl. Soft. Comput.} \textbf{74}, 451-465 (2019).

\bibitem{Kingman1982SPA} J. F. C. Kingman, \emph{Stochastic Process. Appl.} \textbf{13}, 235-248 (1982).

\bibitem{Wakeley2009} J. Wakeley, Coalescent theory: an introduction, (Roberts and Company Publishers, 2009).

\bibitem{Oliveira2012TAMS} R. Oliveira, \emph{Trans. Am. Math. Soc.} \textbf{364}, 2109-2128 (2012).

\bibitem{Allen17Nature} B. Allen, G. Lippner, Y. T. Chen, B. Fotouhi, N. Momeni, S. T. Yau, and M. A. Nowak, \emph{Nature} \textbf{544}, 227-230 (2017).

\bibitem{Szabo1998PRE} G. Szab\'{o} and C. T\H{o}ke, \emph{Phys. Rev. E} \textbf{58}, 69 (1998).

\bibitem{Ohtsuki2006Nature} H. Ohtsuki, C. Hauert, E. Lieberman, and M. A. Nowak, \emph{Nature} \textbf{441}, 502-505 (2006).

\bibitem{Ohtsuki2006JTB} H. Ohtsuki and M. A. Nowak, \emph{J. Theor. Biol.} \textbf{243}, 86-97 (2006).

\bibitem{Ohtsuki2008JTB} H. Ohtsuki and M. A. Nowak, \emph{J. Theor. Biol.} \textbf{251}, 698-707 (2008).

\bibitem{Konno2011JTB} T. Konno, \emph{J. Theor. Biol.} \textbf{269}, 224-233 (2011).

\bibitem{Sample2017JMB} C. Sample and B. Allen, \emph{J. Math. Biol.} \textbf{75}, 1285-1317 (2017).

\bibitem{Chen2013AAP} Y. T. Chen, \emph{Ann. Appl. Probab.} \textbf{23}, 637-664 (2013).

\bibitem{Santos2006PNAS} F. C. Santos, J. M. Pacheco, and T. Lenaerts, \emph{Proc. Natl. Acad. Sci. U.S.A.} \textbf{103}, 3490-3494 (2006).

\bibitem{Fudenberg2006TPB} D. Fudenberg, M. A. Nowak, C. Taylor, and L. A. Imhof, \emph{Theor. Popul. Biol.} \textbf{70}, 352-363 (2006).

\bibitem{Santos2008Nature} F. C. Santos, M. D. Santos, and J. M. Pacheco, \emph{Nature} \textbf{454}, 213-216 (2008).

\bibitem{Santos2012JTB} F. C. Santos, F. L. Pinheiro, T. Lenaerts, and J. M. Pacheco, \emph{J. Theor. Biol.} \textbf{299}, 88-96 (2012).

\bibitem{Allen2014JMB} B. Allen and C. E. Tarnita, \emph{J. Math. Biol.} \textbf{68}, 109-143 (2014).

\bibitem{Cox1989AP} J. T. Cox, \emph{Ann. Appl. Probab.} \textbf{17}, 1333-1366 (1989).

\bibitem{Newman1999PLA} M. E. Newman and D. J. Watts, \emph{Phys. Lett. A} \textbf{263}, 341-346 (1999).

\bibitem{Barabasi1999Science} A. L. Barab\'asi and R. Albert, \emph{Science} \textbf{286}, 509-512 (1999).

\bibitem{Auer15SR} S. Auer, J. Heitzig, U. Kornek, E. Sch\"oll, and J. Kurths, \emph{Sci. Rep.} \textbf{5}, 13386 (2015).

\bibitem{Tucker2018RSO} T. Evans and F. Fu, \emph{R. Soc. Open Sci.} \textbf{5}, 181122 (2018).

\bibitem{Arruda2018PR} G. F. de Arruda, F. A. Rodrigues, and Y. Moreno, \emph{Phys. Rep.} \textbf{756}, 2-59 (2018).

\bibitem{Amaral2020PRE} M. A. Amaral,  W. G. Dantas, and J. J. Arenzon, \emph{Phys. Rev. E} \textbf{101}, 062418 (2020).

\bibitem{Gimenez2021EPJB} M. C. Gimenez, L. Reinaudi, A. P. Paz-Garc\'{\i}a, P. M. Centres, and A. J. Ramirez-Pastor, \emph{Eur. Phys. J. B} \textbf{94}, 1-11 (2021).

\bibitem{Hauser2014JTB} O. P. Hauser, A. Traulsen, and M. A. Nowak. \emph{J. Theor. Biol.} \textbf{343}, 178-185 (2014).

\bibitem{Hindersin2014JRSI} L. Hindersin and A. Traulsen, \emph{J. R. Soc. Interface} \textbf{11}, 20140606 (2014).

\bibitem{Askari2015PRE} M. Askari and K. A. Samani, \emph{Phys. Rev. E} \textbf{92}, 042707 (2015).

\bibitem{Farhang2017PLOSCB} S. Farhang-Sardroodi, A. H. Darooneh, M. Nikbakht, N. L. Komarova, and M. Kohandel, \emph{PLoS Comp. Biol.} \textbf{13}, e1005864 (2017).

\bibitem{Hathcock2019PRE} D. Hathcock and S. H. Strogatz, \emph{Phys. Rev. E} \textbf{100}, 012408 (2019).

\bibitem{Hindersin2019CB} M. M\"{o}ller, L. Hindersin, and A. Traulsen, \emph{Commun. Biol.} \textbf{2}, 137 (2019).

\bibitem{Tkadlec2019CB} J. Tkadlec, A. Pavlogiannis, K. Chatterjee, and M. A. Nowak, \emph{Commun. Biol.} \textbf{2}, 138 (2019).


\bibitem{Sui2015PRE} X. Sui, B. Wu, and L. Wang, \emph{Phys. Rev. E} \textbf{92}, 062124 (2015).

\bibitem{Xiao2019PCB} Y. Xiao and B. Wu, \emph{PLoS Comput. Biol.} \textbf{15}, e1007212 (2019).

\bibitem{Lessard2007JMB} S. Lessard and V. Ladret, \emph{J. Math. Biol.} \textbf{54}, 721-744 (2007).

\bibitem{Maciejewski2014JTB} W. Maciejewski, \emph{J. Theor. Biol.} \textbf{340}, 285-293 (2014).

\bibitem{Tarnita2014AN} C. E. Tarnita and P. D. Taylor, \emph{Am. Nat.} \textbf{184}, 477-488 (2014).

\bibitem{McAvoy2021JMB} A. McAvoy and B. Allen, \emph{J. Math. Biol.} \textbf{82}, 1-41 (2021).


\end{thebibliography}
\end{document}